# Affective responses to chromatic ambient light in a vehicle

Taesu Kim[1], Kyungah Choi[2], Hyeon-Jeong Suk[1]

[1]Department of Industrial Design, Korea Advanced Institute of Science and Technology, Daejeon, Republic of Korea

[2]Department of Interior Architecture Design, Hanyang University, Seoul, Republic of Korea

Corresponding authors:

Kyungah Choi: kchoi@hanyang.ac.kr

Hyeon-Jeong Suk: color@kaist.ac.kr

# Affective responses to chromatic ambient light in a vehicle

This study investigates the emotional responses to the color of vehicle interior lighting using self-assessment and electroencephalography (EEG). The study was divided into two sessions: the first session investigated the potential of ambient lighting colors, and the second session was used to develop in-vehicle lighting color guidelines. Every session included thirty subjects. In the first session, four lighting colors were assessed using seventeen adjectives. As a result, 'Preference, Softness, Brightness, and Uniqueness were found to be the four factors that best characterize the atmospheric properties of interior lighting in vehicles. Ambient illumination, according to EEG data, increased people's arousal and lowered their alpha waves. The following session investigated a wider spectrum of colors using four factors extracted from the previous session. As a result, bluish and purplish lighting colors had the highest preference and uniqueness among ten lighting colors. Green received an intermediate preference and a high uniqueness score. With its great brightness and softness, Neutral White also achieved an intermediate preference rating. Despite receiving a low preference rating, warm colors were considered to be soft. Red was the least preferred color, but its uniqueness and roughness were highly rated. This study is expected to provide a basic theory on emotional lighting guidelines in the vehicle context, providing manufacturers with objective rationale.

# 1. Introduction

Lighting in automobiles has traditionally been used as a visual alarm system to encourage safe driving. More specifically, the role of interior lighting in vehicles was previously described by Löcken et al. (2016) as conveying information regarding the placement of the button, the condition of the car, or the state of the road. However, over the previous few years, lighting technology has improved significantly, making it easier to install and control lighting properties like brightness and color (Blankenbach et al., 2020). Furthermore, as research into how lighting affects users' emotional experiences has become more active (Caberletti et al., 2010, Lu and Fu, 2019, Kim et al., 2021), lighting has begun to be used in a wide range of products. It has also been suggested that lighting may be used in automobiles to influence passengers' emotional experiences. Kia, for example, provided a color-changing mood lighting system inspired by nature to transform vehicle interior into a relaxing and recharging environment (Hyundai Newsroom, 2018). Benz provided ambient lighting with a unique color combination to instill a futuristic mood in the interior, as well as an active ambient light that visually reinforces alerts through color changes (Benz, 2021). As the preceding examples show, lighting color is being actively included by automobile manufacturers as an interior feature; therefore, research into how interior lighting color affects passengers in vehicles should be taken into consideration.

Lighting is one of the complex materials in which user preferences and emotional expectations vary depending on the context of the environment (Choi et al., 2016). Therefore, while there is a growing body of literature on the emotional impact of lighting colors (Yamagishi and Mita, 2014, Odabasioglu and Olgunturk, 2015, Dijck and Heijden, 2005, Westland et al. 2017), it is critical to investigate these effects in specific contexts. In an office, for instance, high color temperature lighting is preferred because it creates a fresh feeling and reduces depression (Bakker et al., 2019). Cool white lighting is also preferrable in a learning environment because it wakes up students (Choi et al., 2019). In residential settings, however, warm white lighting is generally preferred because of its soft mood (Chiazzo et al., 2021). In addition to interior settings, the use of lighting in mass-produced products has also been studied. For example, 8,000 K lighting is preferred for clothing care devices to make clothing look clean and elegant (Kim and Suk, 2020). For refrigerators, customers preferred cool white lighting to make food appear fresher (Raghavan and Narendran, 2002). According to the studies mentioned above, context-specific judgment is essential since expectation and preference to lighting are greatly influenced by the usage situation.

Several research have looked into how colors of lighting affect people's emotions in a vehicle environment. Caberletti and colleagues (2010) proposed 30 ambient illumination settings with two distinct hues, three brightness levels, and five different luminaire arrangements. They found that orange lighting was more comfortable and luxurious than blue lighting, but it was also dark, unimpressive, and less identifiable. Hassib et al. (2019) also evaluated two illumination colors—orange and blue. Regardless of illumination colors, they discovered that ambient lighting improved driver comfort and attentiveness while driving. More recently, Kim et al. (2021) investigated 25 lighting scenarios made up of five colors—red, green, blue, warm white, and cool white—and five animations—blink, rapid blink, dim, rapid dim, and always on. The results showed that red color elicits high arousal while warm white color elicits low arousal. In the earlier studies, the investigations used highly saturated colors, indicating that there is greater color versatility than is typically seen in architectural contexts with a predominance of white illuminations (DiLaura et al. 2011). Furthermore, the illuminance level was limited to less than 4 lx. This shows that the preferred illuminance is lower than typical indoor environments, which is between 300 and 600 lux (DiLaura et al., 2011).

For lighting levels under 100 lx, the visual impact of lighting is typically greater than the non-visual impact (DiLaura et al. 2011). Additionally, when the ambient brightness is 15 times lower than the luminance of the light source, the color of objects that are reflected by the light has a greater impact on human perception than the light source itself (KSA0012). Stylidis et al. (2020) used an online survey to emphasize the quality criteria that customers look for in automotive lighting. They concluded that, unlike headlamps and trunk lighting, automotive interior lighting is directly tied to the interior's luxury, and that uniform illumination of materials is seen as a crucial factor. Lu and Fu (2019) evaluated the impact of ambient vehicle lighting by reflecting 2 lx red, green, blue, cyan, magenta, and yellow lights on interior surfaces. They propose that, in contrast to general lighting evaluation, low illuminance ambient light in a vehicle interior should be evaluated with a focus on how lightings are seen on an automotive interior material.

In summation, this study aims to characterize the atmospheric properties of ambient lighting and assess the effect of colored illuminations in a real vehicle scenario. This study builds on and advances past investigations by widening the range of illumination colors. The study consists of two experiments. The goal of the first study is to identify and extract factors that characterize the atmospheric properties of vehicle interior lighting. By doing so, the study aims to compare the atmospheric expectations for ambient light in a vehicle context with those in other types of spaces. In the second study, diverse lighting colors were examined using the factors extracted from Study 1. The study's ultimate goal is to create a color guideline for vehicle interior lighting by comparing its findings to previous research.

## 2. Study 1: Extracting Factors for Describing the Vehicle Interior Lighting

The objective of this session is to identify the factors that describe the atmospheric qualities of interior illumination in vehicles. Four representative lighting colors were used in the design of the passenger car's ambient lighting (see Section 2.1). We created an evaluation metric for a self-assessment using adjectives gathered from prior studies (see Section 2.2). The electroencephalography was used to measure brain's reactions to ambient lighting (see Section 2.3). Based on the self-assessment and EEG data, we looked at how lighting colors affected drivers' emotional experiences (see Section 2.4).

**2.1 Lighting stimuli**

We used RGBW LEDs (SK6811, Neopixel) to control the color of ambient lighting. We connected individual light sources with light diffusion fiber (3M, USA) to achieve uniform ambient lighting. In order to prevent external color effect, the lighting was fitted in a passenger car with a black exterior and interior (K5, KIA, 2018). When measured by a color meter (CM-2600, Konica Minolta), the color of its seat was $[L^*, a^*, b^*] = [23.18, -0.14, -0.90]$.

The experiment employed Warm White (3,500 K), Neutral White (5,000 K), Cool White (8,000 K), and Purple as the stimuli. Warm White (3,500 K), Neutral White (5,000 K), and Cool White (8,000 K) are the three types of white lighting that are frequently used to illuminate interior spaces. Purple is a color that manufacturers frequently employ in marketing videos to highlight the ambient lighting in their cars (Benz, 2017, Kia, 2018). Figure 1 displays the relative power distribution and chromaticity of each light as measured by the brightness meter (CS-2000A, Konica Minolta).

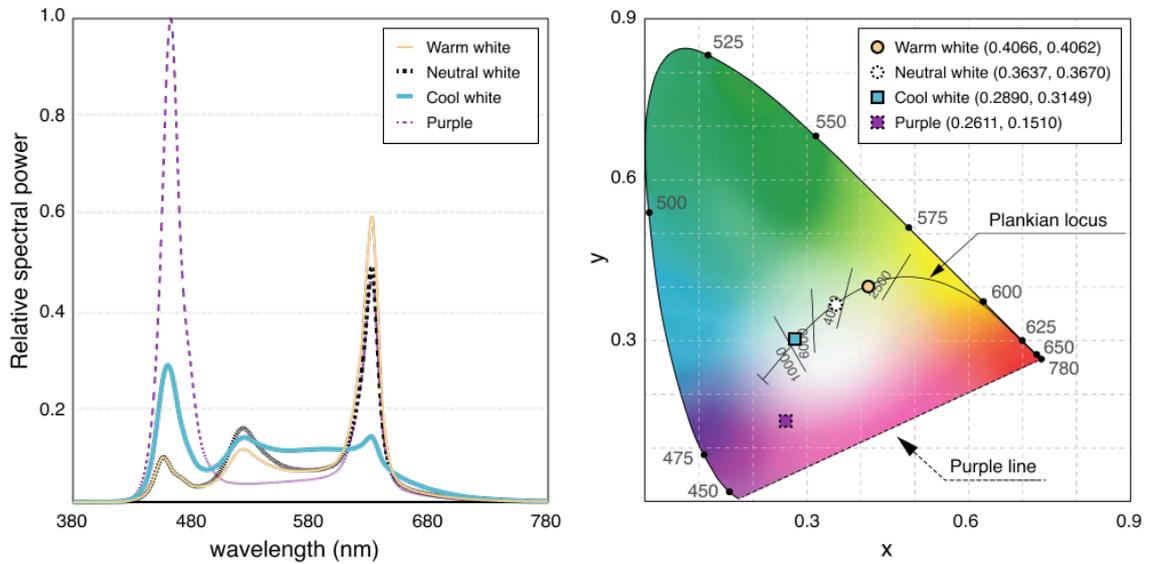

Figure 1. (Left) Relative spectral power of four colors used in the experiment. (Right) The colorimetry value of the lightings plotted in CIE 1931 color space.

We set up three point lights, two edge-lit lights, and one direct-lit light with seven LED chips for our ambient lighting design (Figure 2). According to past research and currently available passenger cars, we decreased its brightness to be under 2 lux at eye level. We parked the car in the subterranean parking lot so that we might fix the nearby illuminants in order to prevent the impact of environmental illumination (5,121 K, 6 lx).

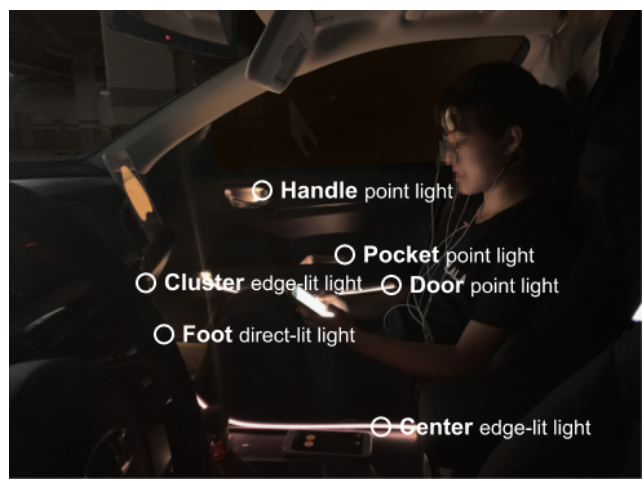

Figure 2. The experiment's lighting configuration. Three point lights (Handle, Pocket, and

Door), two edge-lit lights (Cluster, Centener), and one direct-lit light with seven LED chips (Foot) were used to provide the ambient light. There were fifteen LED chips utilized in all.

**2.2. Assessment method**

2.2.1. Self-assessment

The impact of ambient illumination in vehicles has not been sufficiently studied. Therefore, we pulled adjectives from related fields: 39 adjectives from the research that looked at the effect of illuminants in an architectural context (Flynn et al., 1973; Vogels, 2008; Choi et al., 2015), and 35 adjectives from the subjective assessment of lighting as a visual feedback solution in the vehicle (Caberletti et al., 2010, Huysduynen et al., 2017, Lu and Fu, 2019, Kim et al., 2021). We gathered 74 adjectives, grouped them into 17 groups based on their semantic similarity, and then used those 17 groups to create a self-assessment questionnaire for car ambient lightings. Adjectives utilized in this study included those that describe lighting effects (spacy, bright, calm, clean, soft, warm, tension, pleasant, satisfied), as well as those that describe the style of the vehicle (identifiable, safe, luxurious, trendy, sporty, on-brand, unique, wish to own).

2.2.2. Electroencephalography (EEG)

EEG data is frequently used in studies of how people react emotionally to ambient lighting. In this study, five electrodes were placed on the subject in order to monitor brain activity: Fp1, Fp2, ground potential, and two reference potentials (A0, A1) (Figure 3). Data was gathered at 250 Hz using the signal gathering device (BIOS-S24, Biobrain).

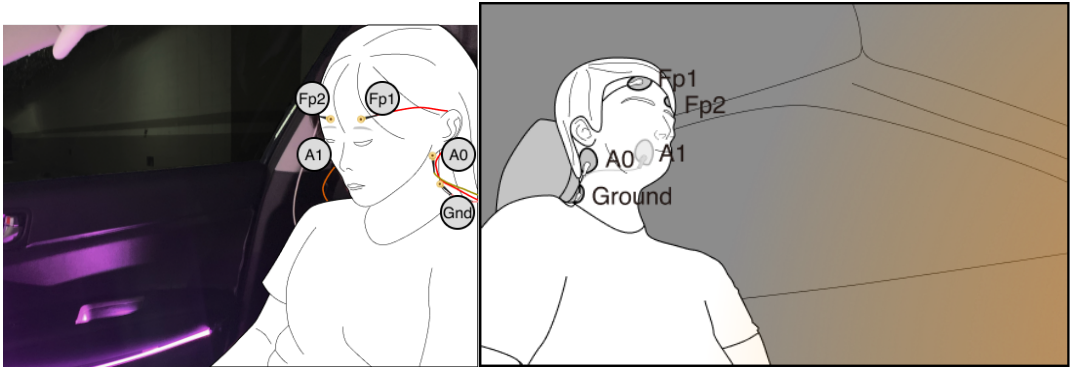

Figure 3. The experimental setup for electroencephalography. Five points were gathered utilizing the device: Fp1, Fp2, Ground signals, A0, and A1 (BIOS-S24, Biobrain)

**2.3. Procedure of the experiment**

A total of 30 participants (M = 22.5, SD = 2.9) with at least a year of driving experience and sensitive to trends with interest in the car took part in the study. 15 men and 15 women were selected as volunteers, with the gender ratio being controlled to avoid gender effects. The participants had to pass the Ishihara test to demonstrate that none of the individuals had color blindness. They received $15 as payment following the experiment.

The experiment was conducted with subjects seated in the car's seat. While the lights were off, we connected the electrode and recorded their baseline state for a minute. Then, we offered four different lighting colors in a randomized order. Prior to evaluation, we asked them to take a moment to scan the illumination. After they had seen the illumination, we asked them to complete the self-assessment forms provided in a 5-Likert Scale (with −2 being strongly disagree, and 2 being strongly agree). We then collected their EEG for a minute. Before moving to a new lighting color, we advised participants to close their eyes for a minute to offset the effects of the prior lighting color. At the end of the experiment, we reassessed the ground condition of participants. The experiment was completed in 45 minutes.

## 2.4. Results

The factor analysis was performed to extract the adjectives for describing the atmospheric characteristics of ambient lighting in the vehicle. The sampling adequacy value calculated by Kaiser-Meyer-Olkin test was 0.85, and Bartlett's test of sphericity had a significant result ($\chi^2(136) = 1016.55$, $p < .05$). The principal component analysis using the Varimax rotation approach was used for the factor analysis on all seventeen adjectives.

As a result, seventeen adjectives were grouped into four factors. The four factors that were extracted were "Prefer - Not prefer," "Soft - Rough," "Bright - Dark," and "Unique - Common." The results of the factor analysis are presented in Table 1.

Table 1. The factor loadings of the four factors (KMO = .85, Bartlett's significance value < .05).

|  | Not prefer - Prefer (Eigenvalues 4.62, Variance 27.16%) | Rough - Soft (Eigenvalues 2.70, Variance 15.86%) | Darkness - Brightness (Eigenvalues 2.23, Variance 13.13 %) | Common - Unique (Eigenvalues 1.57, Variance 9.24 %) |
|---|---|---|---|---|
| satisfied | **.792** | .282 | .282 | .076 |
| wish | **.784** | .234 | .230 | .100 |
| trendy | **.767** | -.028 | .272 | .369 |
| luxurious | **.731** | .121 | .163 | .315 |
| consistent | **.728** | .144 | .302 | -.045 |
| comfortable | **.622** | -.034 | .569 | -.096 |
| sporty | **.606** | -.416 | .338 | .258 |
| identifiable | **.597** | -.048 | .025 | .044 |
| calm | **.492** | .412 | -.221 | -.430 |
| warm | -.227 | **.831** | -.153 | .089 |
| soft | .193 | **.777** | -.006 | -.118 |
| pleasant | .237 | **.653** | .304 | .104 |
| safe | .201 | **.544** | .277 | -.156 |
| brightness | .205 | .073 | **.838** | -.020 |
| spacy | .246 | .055 | **.717** | .115 |
| unique | .218 | .122 | -.066 | **.837** |
| tension | .186 | -.373 | .083 | **.533** |

The effect of light chromaticities was investigated using a repeated-measures one-way ANOVA for the loading scores of each of the four components. There were statistically significant differences between the lightings across all four components ($F_{preference}$ [3,87] = 26.73, $p < .05$, $F_{softness}$ [3,87] = 17.02, $p < .05$, $F_{brightness}$ [3,87] = 70.20, $p < .05$, $F_{uniquness}$ [3,87] = 74.10, $p < .05$). The results are shown in Figure 4.

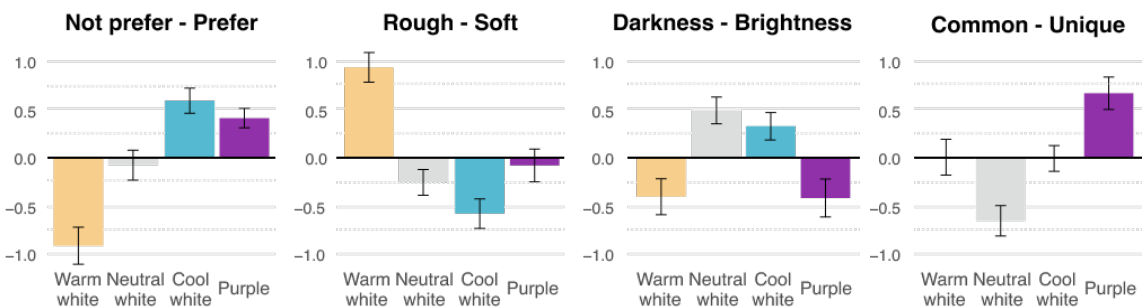

Figure 4. Average ratings for each of the four lighting colors in terms of preference, softness, brightness, and uniqueness. Error bars indicate ± Standard Error

In terms of preference, participants favored Cool White and Purple lighting but did not favor Warm White lighting. Cool White lighting was rated as *rough* and *bright* illumination. Purple lighting gave the car a *unique* ambience, although it was seen as being *darker* than other colors. Subjects described the Warm White as *soft* and *dark*. Neutral White made the car *bright* yet was seen as *common* lighting for a car.

In order to determine the impact of ambient lighting on brain responses, a paired sample t-test was performed to compare ambient lighting with the condition where it was turned off (see Figure 5). The alpha wave was impacted by chromatic illumination, which resulted in a lower alpha wave in subjects when compared to the condition where the lights were off (one-tailed, $p < .05$). There were no statistically significant differences in beta waves. Repeated measures one-way ANOVA was used to compare four lighting colors, however the results showed no significant differences ($F_{Alpha}[3,87] = 1.92$, $p = .13$, $F_{Beta}[3,87] = .46$, $p = .71$).

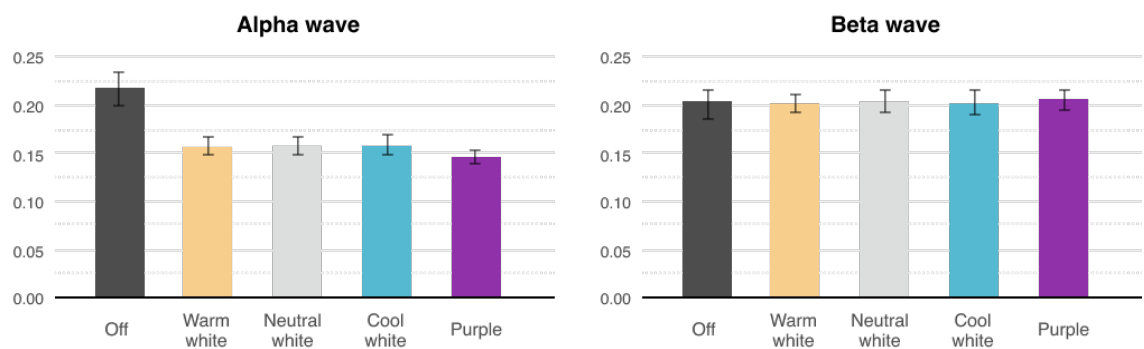

Figure 5. Alpha wave ratio and beta wave ratio the four lightings. Error bars indicate ± Standard Error

## 2.5. Discussion

In this study, we investigated the impact of lighting colors on how passengers perceive the ambiance in a car. We created four chromaticities of ambient lighting for the passenger automobile and used factor analysis to obtain adjectives that described the impact of vehicle interior lighting. We used the extracted factors to compare the self-assessment findings in order to define the impact of the lighting chromaticities. The effects of brain responses were then evaluated by examining the EEG alpha and beta waves.

According to the results of the factor analysis, four factors—*prefer*, *soft*, *bright*, and *unique*—were extracted. Comparing our results to Flynn's research, a well-known study for subjective evaluation of ambient lighting, the criteria *general impression* and *visual clarity* were also extracted in our experiment. They were in accordance with the *preference* and *brightness* factors that were taken from our study, respectively (Flynn et al. 1973). The *soft* factor is consistent with Vogel's research, which recommended that *coziness* and *liveliness* should be included when assessing ambient lighting in the setting of a room (Vogels, 2008). The *spaciousness* factor suggested by Vogel, however, was not extracted from this study.

The *uniqueness* factor was extracted from the factor analysis, which is a term that isn't usually used to evaluate ambient illumination at architectural settings. Vehicles, unlike interior spaces, are mass-produced. Consumers judge how unique a product is in comparison to others when purchasing mass-produced goods, and this tendency is particularly noticeable when purchasing vehicles (Zaggl et al. 2019). Given that the focus of this experiment was in-vehicle scenarios, the emotional assessment of the interior illumination may have also been influenced by the vehicle purchasing criterion. In recent years, lighting has become almost as prevalent in mass-produced items as interior spaces. Therefore, when evaluating the user's perception to illumination, the *uniqueness* criterion should be taken into consideration.

According to the study, The *preference* score for Cool White was the highest. Previous studies have shown that Cool White promotes concentration because it helps people perceive space as being more bright and rough (Bakker et al., 2019). We assumed that given the driving context, Cool White lighting is preferred for ambient lighting in a car because drivers need to pay close attention when driving. Warm White, on the other hand, received the lowest rating for *preference*, which is inconsistent with how ambient lighting is typically preferred for indoor spaces. Warm White lighting received the highest score for *softness*, in keeping with earlier lighting research. The study found that Warm White correlated with a *soft*, *dark*, and *common* ambience, which may have been related to a lower *preference* score because drivers should be concentrating on the road. Interestingly, despite having a lower *brightness* rating, Purple light demonstrated strong *uniqueness* and *preference*. On the other hand, Neutral White illumination received a low *preference* rating, indicating chromatic lighting's potential as vehicle interior lighting.

Furthermore, we found that ambient illumination decreased the alpha wave ratio of the subjects, suggesting that ambient lighting in a car helps alert drivers. Canazei et al. (2021) also found that bright light could decrease drivers' alpha spindles, indicating that bright ambient light can alert drivers. Because alertness is one of the primary functions of the visual system in the car, our research showed that ambient illumination can have a substantial impact in the vehicle. The investigation did not, however, reveal any differences in chromaticities.

## 3. Study 2: Effect of Chromaticities in Vehicle Interior Lighting

This chapter examines how various lighting chromaticities impact drivers' emotions. A wider spectrum of colors were investigated in order to produce color recommendations for ambient lighting in automobiles. For the experiment, we first constructed the ambient lighting and chose 10 lighting colors (Section 3.1). After that, we collected participants' brain waves while asking them to self-evaluate for four factors that were taken from the preceding study (Section 3.2). Then, we conducted data analysis to compile information on how the colors used in the vehicle's ambient lighting affect people's emotions (Section 3.3).

### 3.1. Lighting Stimuli

The RGBW LEDs (SK6811, Neopixel) and light diffusion fiber (3M, USA) utilized in the preceding part were also employed in the lighting design. We installed the lights on the same passenger car (K5, KIA, 2018) as before.

We conducted the experiment using a total of ten colors. The entire color spectrum was made up of six chromatic lightings and four white lightings. The chromatic lightings are composed of three primary colors—Red, Green, and Blue—as well as three additional colors—

Orange, Yellow, and Purple. We used color presets from the KIA K900 and the Benz S class, two well-known manufacturers of passenger automobiles that make use of ambient lighting. For white lightings, we selected 3,000 K, a warm color, 5,000 K and 6,500 K, both of commonly used for neutral colors, and 10,000 K, which is a representation of cool white. Figure 6 and Table 2 displays the chromaticity of each light as measured by the luminance meter (CS-2000A, Konica Minolta).

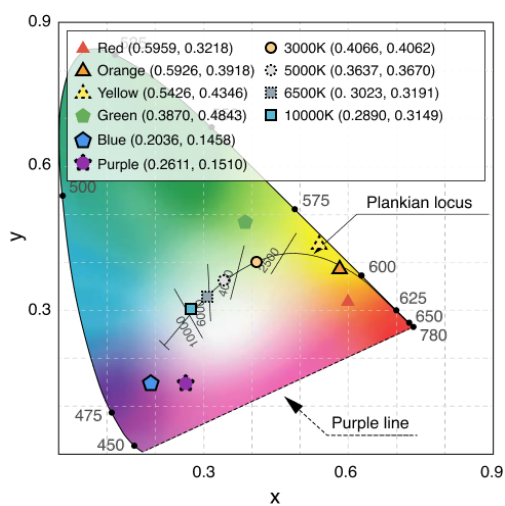

Figure 6. The colorimetry value of the lightings plotted in the CIE 1931 color space.

Table 2. The CIE1931 *xy* value and dominant wavelength of the lightings used in the experiment.

| color | *x* | *y* | dominant wavelength (nm) |
|---|---|---|---|
| Red | 0.5959 | 0.3218 | 615.7 |
| Orange | 0.5926 | 0.3918 | 595.4 |
| Yellow | 0.5426 | 0.4346 | 587.1 |
| Green | 0.387 | 0.4843 | 566.6 |
| Blue | 0.2036 | 0.1458 | 466.3 |
| Purple | 0.2611 | 0.1510 | -567.7 |
| 3,000 K | 0.4066 | 0.4062 | 578.1 |
| 5,000 K | 0.3637 | 0.3670 | 576.8 |
| 6,500 K | 0.3023 | 0.3191 | 485.6 |
| 10,000 K | 0.2890 | 0.3149 | 486.2 |

Instead of the passenger seat, which was employed in earlier experiment, the lights were mounted on the back seat. We adjusted the ambient lighting so that it was 2 lux or less at the eye level. The car was parked underground to control the environment illuminant (5,121 K, 6 lx), as shown in Figure 7. This helped us avoid the effects of the environmental illuminants.

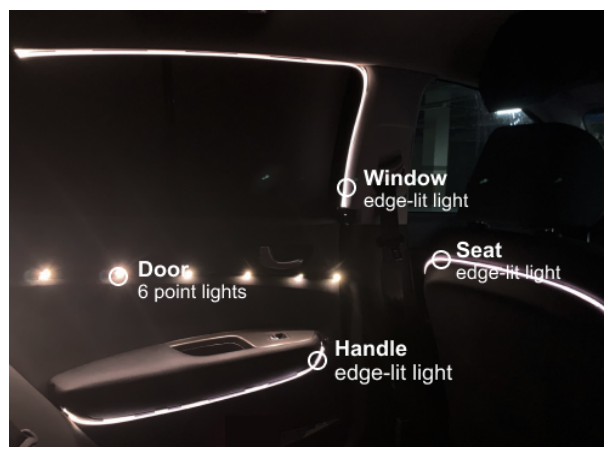

Figure 7. The experiment's lighting configuration. Six point lights (along the door trim) and three edge-lit lights (Window, Seat, Handle) were used to provide the ambient light. There were twelve LED chips utilized in all.

### 3.2. Assessment method

3.2.1. Procedure

For user evaluation, we gathered survey and brain activity data from participants. The questionnaire was constructed using four factors—*Not prefer-Prefer, Rough-Soft, Darkness-*

*Brightness, Common-Unique*—extracted from the preceding session. The survey was provided in a 5-Likert Scale. As shown in Figure 3, we recorded brain activity using an EEG from the frontal cortex.

The ground state EEG signal of the subject was recorded for a minute. After that, we offered ten colors in a random order. We recorded the EEG of the participants for a minute throughout the light exposure. The participants were then asked to complete the questionnaire. We exposed participants to the light again for one minute after the survey was finished in order to record EEG once more. We measured the ground condition again for a minute after presenting ten lighting colors. It took 50 minutes to complete the experiment.

3.2.2. Participants

The study included 30 participants (M = 22.0, SD = 2.4) who had at least one year of driving experience, were trend-sensitive, and were interested in cars. In order to prevent gender effects, participants were chosen at a ratio of 15 men to 15 women. To prove that none of the participants had color blindness, the volunteers had to pass the Ishihara test. After finishing the experiment, they were paid $20.

### 3.3. Results

To determine the impact of light colors on each factor, a repeated measures one-way ANOVA was conducted. Significant differences were observed across all variables ($F_{preference}$ [9,261] = 5.14, $p < .05$, $F_{softness}$ [9,261] = 15.10, $p < .05$, $F_{brightness}$ [9,261] = 2.10, $p < .05$, $F_{uniqueness}$ [9,261] = 12.16, $p < .05$). The results are shown in Figure 8.

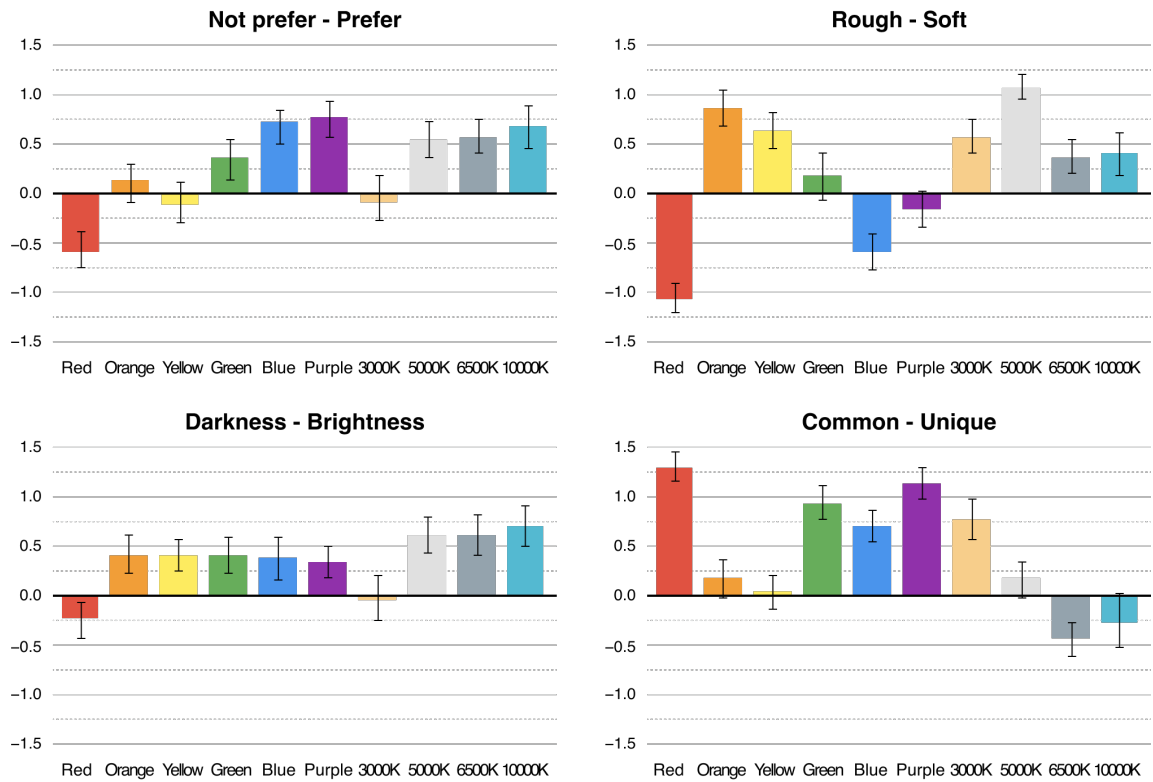

Figure 8. Average ratings for each of the ten lighting colors in terms of preference, softness, brightness, and uniqueness. Error bars indicate ± Standard Error

The experiment's findings indicated that the highest preference was for bluish colors—Blue, Purple and 10,000 K. Following that, Green, 5,000 K, and 6,500 K were preferred. On the other hand, orangish colors—Orange, Yellow, and 3,000 K—were not preferred. The color Red had the lowest preference score.

Purple and Blue chromatic lightings received high ratings for *preferences* and *uniqueness*, but not for *brightness*. These illumination also received a low *softness* rating. Cool White, the whitish lighting that received a high preference rating, was rated highly for being both *soft* and *bright* but was also rated as a *common* color.

Green scored highly on *uniqueness* among middle-ranked ambient lighting colors, but its *softness* and *brightness* scores were relatively low. The 5,000 K and 6,500 K, whose preferences were comparable, were evaluated as *bright* and *soft*, but their *uniqueness* was ranked lower.

Lower preference colors like Orange, Yellow, and 3,000 K were found to create a *soft* ambiance but their brightness was rated as being relatively *dark*. Among these three lightings, the chromatic Orange and Yellow lights were classified as *common* illumination. Red lighting, which received the lowest preference rating, was thought to be *unique* but *rough* and *dark*.

Next, a paired sample t-test was used to compare the brain responses to ambient lightings to the condition when the lights were off. The findings are shown in Figure 9. In comparison to the condition when the lights were off, drivers' alpha waves were lowered when lighting was present, regardless of color (one-tailed, $p < .05$). There were no significant differences observed in beta waves. In order to examine the effects of colors, a repeated measures one-way ANOVA was also performed, but no differences between colors were found to be statistically significant ($F_{Alpha}[9,261] = 1.17$, $p = .32$, $F_{Beta}[9,261] = .94$, $p = .49$).

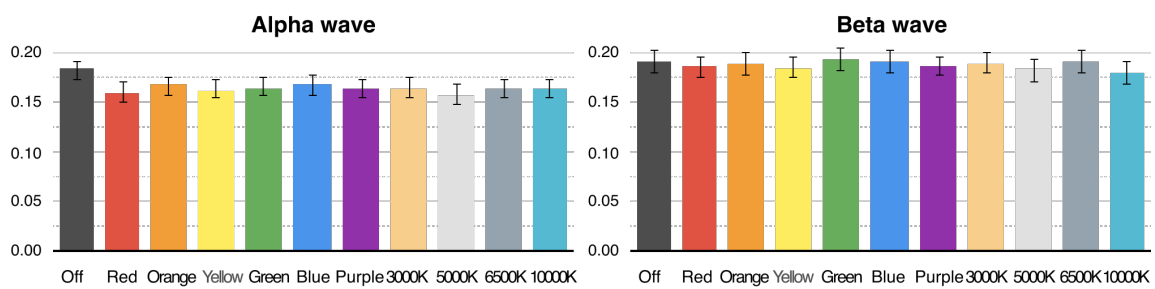

Figure 9. Alpha wave ratio and beta wave ratio the ten lightings. Error bars indicate ± Standard Error

## 3.4. Discussion

This chapter presented results of an investigation of the influence of color on vehicle ambient lighting. Five additional chromatic lightings—Red, Orange, Yellow, Green, and Blue—were added to Purple, a chromatic illumination examined in Study 1. We also chose 3,000 K, a warm color, 5,000 K and 6,500 K, both frequently used for neutral colors, and 10,000

K, a depiction of Cool White, for white illumination.

Regarding *preference* and *brightness*, the findings from this study are consistent with those from Study 1. However, it is noteworthy to note that overall, this study's ratings were higher than Study 1's evaluations. It is anticipated that Red lighting, which obtained the lowest ratings for *preference* and *brightness*, would have a relative influence on how other stimuli were assessed, leading to higher scores.

When compared to Study 1, *softness*, however, demonstrated a distinct tendency. It was confirmed that the 5,000 K and 10,000 K scores were higher than in Study 1, whereas the Warm White score decreased. Yamagishi and Mita (2014) discovered that bluish color lighting with lower saturation and orange color lighting with higher saturation can boost the impression of softness in a room. In contrast to Study 1, we used Orange and Blue color illumination for the experiment, which may have an impact on how the evaluation results alter. It is expected that Blue lighting will have a relative influence on how Cool White (10,000 K) lighting is evaluated. Also, because of the Orange lighting, the *softness* of Warm White (3,000 K) with low saturation may have been perceived as relatively low.

The results from *uniqueness* also differed from those of Study 1. In Study 1, the *uniqueness* of 5,000 K and 10,000 K was comparable. However, according to this study, 5,000 K lighting had a higher level of *uniqueness*. The investigation was conducted on Korean consumers who typically illuminate their spaces with 6,500 K lights. Therefore, 6,500 K neutral lighting would have a relative impact on how 5,000 K neutral lighting was evaluated (Cho, 2021).

The results of the EEG data were comparable to those of Study 1. Vehicle ambient light reduced subjects' alpha wave ratio regardless of color, and no significant differences in beta weves were observed.

In conclusion, although the results from Studies 1 and 2 differed slightly, this could have been because of the stimuli's composition. As a result, we will outline color guidelines based on the findings from Study 2 and discuss their limits as well as our future research goals.

## 4. General Discussion

In this study, an experimental study was carried out using a survey and an EEG to examine the effects of ambient lighting color in the context of a vehicle. In the first session, an experimental study was presented using four colors to examine the possibility of ambient illumination colors influencing the user's emotion. In order to create color recommendations for the vehicle ambient lighting, the next session looked into the impact of ten illuminant colors.

### 4.1. Color guidelines for vehicle ambient lighting

Based on the preference data from Study 2, we proposed guidelines for the color of car ambient lighting. According to the findings, the three most popular colors were Blue, Purple, and 10,000 K. Previous research also revealed that when it came to the color of ambient lighting in vehicles, drivers preferred blue to orange or red (Caberletti et al., 2010; Kim et al., 2021). More specifically, 10,000 K should be used instead of Blue or Purple to create a *soft* and *bright* ambiance inside the vehicle. Previous research has also found that lowering the saturation of lighting may provide a softer ambience (Yamagishi and Mita, 2014, Jin et al., 2015). However, if the vehicle is intended to be more *unique*, the color Purple should be used for ambient lighting.

Green, 5,000 K, and 6,500 K were the next most preferred lighting colors. Neutral White colors—5,000 K and 6,500 K—might have been preferred since these are common colors for indoor lighting. Comparatively, prior studies have indicated that Green is inappropriate for indoor spaces. When used as room illumination, the human eye is sensitive to the wavelength of the color Green, which is uncomfortable (Ayama and Ikeda, 1998, Odabasioglu and Olgunturk, 2015). Green illumination appears to be preferred in this investigation due to the low brightness level of the lighting in the car. To be more specific, it is recommended for manufacturers to use 5,000 K lighting if one of these three colors is used to create a *soft* ambiance in the interior of the vehicle. When attempting to be *unique*, though, manufacturers are encouraged to employ Green illumination.

Low preferences were given to 3,000 K, Yellow, and Orange. We suggest utilizing Orange and Yellow from these colors if the interior of the vehicle needs a slightly *brighter* atmosphere. However, we suggest using 3,000 K light if the vehicle needs a *unique* ambiance. The preference for Red lighting was the lowest. Red was viewed as being *unique*, yet being *rough* and *dark*. Red light has mostly been employed as a warning in vehicles (Dijck and Heijden, 2005), hence it is not appropriate for interior illumination that is always on (Odabasioglu and Olgunturk, 2015).

The study discovered that lighting, regardless of color, reduces the alpha waves of the drivers. Alpha waves decrease when people are engaged in an attention-related task (Ray and Cole, 1985). This demonstrates that turning on a vehicle light can cause drivers to pay more attention. Existing research has also demonstrated that alpha waves are decreased by the lighting in the vehicle (Canazei et al., 2021). The investigation did not, however, reveal any differences in chromaticities. The brightness of ambient light was significantly lower than in earlier studies that examined the impact of light chromaticities, which showed that the impact of colors should not be sufficient to alter brain activity. In conclusion, the use of in-vehicle lighting, regardless of color, encourages drivers to pay closer attention. Beyond this function, however, the use of various lighting colors may contribute to the creation of an atmosphere based on emotional demands.

**4.2. Limitations and future works**

The limitations of this study are as follows. First, different luminaire configurations were used for Studies 1 and 2. Previous research has found that luminaire arrangement influences people's affective perception (Caberletti et al., 2010, Locken et al., 2013), and it is possible that this was the case in this study as well. As the final color guideline was derived from Study 2, further study using different luminaire arrangements is also required.

Next, due to safety concerns, the study was conducted in a parking lot with a static state. However, the experience of driving and non-driving is distinct, and previous research has attempted to simulate the driving condition using simulators (Caberletti et al., 2010, Lu and Fu, 2019, Hassib et al., 2019). Therefore, for the lighting to be embedded in the actual vehicle, a study in a real driving context is also required.

For the experiment, only Korean drivers were surveyed. The research was conducted in a real car, making it challenging to run trials with individuals from different cultures. Weirich et al. (2022) conducted a cross-cultural study of vehicle illumination using an online survey and found that culture affects how people perceive the lighting. Therefore, a study based on the targeted culture or a cross-validation of users from different countries is required in order to adopt the color guidelines recommended by this study.

Next, the experiment was carried out using a single color without using other color combinations. The degree of color flexibility in this experiment was constrained by the use of edge-lit lighting. However, as technology advances, the level of flexibility offered by lighting increases (Blankenbach et al., 2020). Therefore, future considerations for elements such as color combination or lighting color gradation patterns are required.

In this experiment, we used low brightness lighting in accordance with vehicle regulations. Lighting regulations may become more flexible as autonomous vehicle technology advances. Therefore, non-visual effects on drivers' physiological systems should be investigated in the future for high-intensity lighting.

## 5. Conclusion

This study investigated the color effects of ambient lighting in vehicles. Self-assessment and EEG were used to examine how vehicle interior lighting color may affect users. The study consisted of two parts, and it initially explored the possible effects of colored ambient lighting. Subsequently, lighting color guidelines were developed in the second part based on an evaluation of the effects of 10 colors. According to the first part, *preference*, *softness*, *brightness*, and *uniqueness* were the four criteria that describe the impact of ambient lighting color in cars. Additionally, the EEG data showed that ambient lighting made the individuals in the vehicle more alert. The color guideline for ambient lighting was developed in the second part. Specifically, bluish lightings, such as Blue, Purple, and 10,000 K, were the most preferred among the 10 lighting colors, followed by Green and Neutral White lighting (6,500 K and 5,000 K). Warm-colored lightings, such as Orange, Yellow, and 3,000 K, were less preferred. The color Red demonstrated the lowest preference for ambient lighting. These findings suggest that existing lighting knowledge regarding the effects of ambient lighting must be modified when applied to the vehicle context. We anticipate that the findings of this study will provide automobile manufacturers with basic knowledge about the effects of ambient lighting colors.

## Acknowledgments

This research was supported by the 4th BK21 through the National Research Foundation of Korea (NRF) funded by the Ministry of Education (MOE) (NO.4120200913).

The authors report there are no competing interests to declare.

Figure 1. (Left) The relative spectral power of the four colors used in the experiment. (Right) The colorimetry value of the lightings plotted in CIE 1931 color space.

Figure 2. The experimental lighting configuration. Three point lights (Handle, Pocket, and Door), two edge-lit lights (Cluster, Center), and one direct-lit light with seven LED chips (Foot) were used to provide the ambient light. In total, 15 LED chips were utilized.

Figure 3. The experimental setup for electroencephalography. Five points were gathered utilizing the device: Fp1, Fp2, ground potential, A0, and A1 (BIOS-S24, Biobrain)

Figure 4. Average ratings for each of the four lighting colors in terms of preference, softness, brightness, and uniqueness. Error bars indicate ± standard error.

Figure 5. Alpha wave ratio and beta wave ratio of the four lightings. Error bars indicate ± standard error.

Figure 6. The colorimetry value of the lightings plotted in the CIE 1931 color space.

Figure 7. The experimental lighting configuration. Six point lights (along the door trim) and three edge-lit lights (Window, Seat, Handle) were used to provide the ambient light. In total, 12 LED chips were utilized.

Figure 8. Average ratings for each of the 10 lighting colors in terms of preference, softness, brightness, and uniqueness. Error bars indicate ± standard error.

Figure 9. Alpha wave ratio and beta wave ratio the 10 lightings. Error bars indicate ± standard error.

Table 1. The factor loadings of the four factors (KMO = .85, Bartlett's significance value < .05).

Table 2. The CIE1931 *xy* value and dominant wavelength of the lightings used in the experiment.